\documentclass[UKenglish]{article}
\usepackage{epsfig,graphicx,amsmath,amssymb,color,subfigure,url}
\usepackage{natbib} 
\usepackage[UKenglish]{babel}
\usepackage[a4paper,left=30mm,right=25mm]{geometry}

\title{Evidence of bias in the Eurovision song contest: modelling the votes using Bayesian hierarchical models}
\author{Marta Blangiardo$^{1}$ \and  Gianluca Baio$^2$}
\date{
\begin{small}
  $^1$\textit{Department of Epidemiology and Biostatistics, Imperial College London (UK)} \\[.2cm]  
  $^2$\textit{Department of Statistical Science, University College London (UK)} \\[.2cm]
  \today
\end{small}
}
\begin{document}
\setlength{\baselineskip}{18pt}
\maketitle

\vspace{20pt}\noindent\textbf{Running title:} Bayesian modelling for the votes of the Eurovision contest 

\vfill 
\noindent For correspondence, please contact:

\vspace{5pt}\noindent Gianluca Baio, PhD, University College London, Department of Statistical Science, Gower Street, London, WC1E 6BT (UK).

\noindent Email address: \texttt{gianluca@stats.ucl.ac.uk} \newpage

\newpage

{\fontsize{16}{22} \selectfont \begin{center}\begin{minipage}[c]{16cm}
\centering Evidence of bias in the Eurovision song contest: modelling the votes using Bayesian hierarchical models
\end{minipage}\end{center}}

\vspace{40pt}

\begin{center}
{\fontsize{14}{16}\selectfont \textbf{Abstract}}
\vspace{20pt}

\begin{minipage}[c]{13cm}
The Eurovision Song Contest is an annual musical competition held among active members of the European Broadcasting Union since 1956. The event is televised live across Europe. Each participating country presents a song and receive a vote based on a combination of tele-voting and jury. Over the years, this has led to speculations of tactical voting, discriminating against some participants and thus inducing bias in the final results. In this paper we investigate the presence of positive or negative bias (which may roughly indicate favouritisms or discrimination) in the votes based on geographical proximity, migration and cultural characteristics of the participating countries through a Bayesian hierarchical model. Our analysis found no evidence of negative bias, although mild positive bias does seem to emerge systematically, linking voters to performers.

\vspace{10pt}
\end{minipage}
\end{center}

\newpage

\section{Introduction}
The Eurovision Song Contest is an annual musical competition held among active members of the European Broadcasting Union (EBU). The first edition of the contest was held in 1956 in Lugano (Switzerland). The event was televised live across Europe, in what represented a highly technological experiment in broadcasting.

Members of the EBU approved plans to hold the contest on an annual basis and there were initially seven participating countries: The Netherlands, Switzerland, Belgium, Germany, France, Luxembourg and Italy, each entering the competition with two songs. The winner was decided by a jury which consisted of an equal number of members per participating country.

The voting system in the contest has changed over time. From 1962 onwards, positional voting was used, eventually leading to the current point system, in which 12, 10, 8, 7, 6, 5, 4, 3, 2, 1 points are allocated to each country's top 10 favourite songs. The country with the highest score overall is announced as the winner.

Tele-voting was introduced in 1997 and allowed viewers from participating nations to vote for their favourite act via phone, email or text. Austria, Germany, Sweden, Switzerland and the United Kingdom trialled the system, while the rest continued using juries. In 1998 all countries used tele-voting to determine the points awarded to the top ten preferred acts, and from then onwards all countries have used this system or a mixture of tele-voting/juries to determine the way in which points are allocated.

Especially with the introduction of tele-voting, accusations of bias in the voting system have been brought forward by several commentators. Famously, in 2008 Sir Terry Wogan announced that he would quit as the BBC's Eurovision contest commentator after casting doubts over the regularity of the contest \citep{BBC:2008}. Periodically, the media investigate accusations of wrong-doing in the management of the contest \citep{BBC:2012} and the problem of bias and political influence over the voting system of the Eurovision contest has been also considered in the scientific literature.

\citet{Yair:1995} is probably the first paper addressing the issue of collusive voting behaviour in the contest; his analysis based on multidimensional social networks showed the presence of three main ``bloc'' areas: Western, Mediterranean and Northern, although no detailed statistical assessment was given of the derived associations among countries attitude towards each other. \citet{Clerides:2006} used an econometric model to quantify the impact of factors determining \textit{affinity} and \textit{objective quality} on the actual votes. Their conclusions were that some evidence of reciprocity was found, but no strategic voting resulted from the analysis. \citet{Fenn:2006} used dynamical network and cluster models to show that while the existence of ``unofficial cliques'' of countries is supported by the empirical evidence, the underlying mechanism for this cannot be fully explained by geographical proximity.  \citet{Spierdijk:2006} investigated how geographical, cultural, linguistic, and religious factors lead to voting bias using multilevel models and considering the bias of one country towards another as the dependent variable. Their analysis points to evidence to suggest that geographical and social factors influence certain countries voting behaviour, although political factors did not seem to play a role in influencing voting. In a similar vein, \citet{GinsburghNoury:2008} argue that determinants other than political conflicts or friendships, such as linguistic and cultural proximity, seem to be mostly associated with the observed voting patterns. 

All in all, the existing scientific evidence seems to suggest that indeed there are particular voting patterns that tend to show up more often than not; however, it is less clear whether this can be taken as definitive proof of the existing of fundamental bias, either in terms of favouritism or discrimination. In this paper we aim at quantifying the presence of systematic bias in the propensity to vote for a given performer. We use a Bayesian hierarchical framework to model the score as a function of a random (structured) effect which depends on \textit{cultural} and \textit{spatial} proximity, as well as on \textit{migration} stocks. Using this strategy we aim at capturing the possible effects of social as well as geographical components which might influence the voting patterns. Moreover, we control for some potential confounder factors, \textit{i.e.} the year in which the contest was held, the country hosting the contest, the language in which each song was sung and the type of act (male solo artist, female solo artist or mixed group).

As we will discuss later, we are not particularly interested in the ``effect'' of these covariates on the score associated with a given voter, a given performer and a given occasion. Rather, we use these to balance the data and account for potentially different baseline characteristics. Nor are we focussed on predicting the actual votes for next instance of the contest, given them. The main objective of the paper is to try and identify the impact of the social and geographical structured effect on the voting patterns and thus, unlike many regression models, the interest of our analysis lies almost exclusively on the random effects.

The rest of the paper is structured as follows: Section \ref{Data} presents the available data and the variables used in the model; Section \ref{BayModel} specifies the Bayesian framework used for the analysis including the model fit index used to find the best specification; Section \ref{Results} presents the results for the best-fitting model, and finally Section \ref{Discussion} discusses some issues related to the model.

\section{Data}\label{Data}
In this analysis we use data on the final round of votes of the contest during the period 1998-2012 inclusive. This period is selected for pragmatic reasons, since tele-voting was only adopted from 1998 onwards. The data are available from the official Eurovision contest website (\texttt{www.eurovision.tv}).

All countries that have voted in the final round in the period under study have been considered in our analysis. For each combination of voter, performer and year, the votes are available as an ordinal categorical variable, which can assume values \{0,1,2,3,4,5,6,7,8,10,12\}.

The available predictors are the following: the language in which each song was sung (the performer's language, English, or a mixture of two or more languages), the gender and the type of performance (group, solo male artist, solo female artist). We specify the random effects as a function of data on two dimensions: first we consider the migration stocks, obtained from the World Bank's dataset (\texttt{www.worldbank.org}) as a proxy of the migration intensity from the voter's to the performer's country. This is supposed to account for possible favouritism in voting patterns due to the presence of large stocks of people originally from the performer's country, but currently living in the voter's country. Secondly, we consider the neighbouring structure, defined in terms of the countries sharing boundaries. This is used to account for similar geographical characteristics.

\section{Bayesian modelling}\label{BayModel}
We define the voters as $v=1,\ldots,V=48$ and the performers as $p=1,\ldots,P=43$ (\textit{i.e.} our data contain some countries that vote but do not perform). The outcome of interest is the variable $y_{vpt}$ representing the points given by voter $v$ to performer $p$ on occasion (year) $t=1,\ldots,T_{vp}$. Thus, $y_{vpt}$ is a categorical variable which can take any of the $S=11$ values in the set of scores $\mathcal{S}=\{0,1,2,3,4,5,6,7,8,10,12\}$. Note that the number of occasions for the voter-performer pair ($T_{vp}$) can vary between $0$ and $15$ in the dataset considered. Moreover, because not all the countries have participated consistently throughout the several editions of the contest, the dataset is not balanced and therefore the value $T_{vp}$ does vary with the pair $(v,p)$. In particular, this means that there are $H=1937$ observed combinations of voter-performer pairs.

We then model 
\begin{eqnarray*}\label{Likelihood}
y_{vpt} \sim \mbox{Categorical}(\boldsymbol{\pi}_{vpt}),
\end{eqnarray*}
where $\boldsymbol{\pi}_{vpt}=(\pi_{vpt1},\ldots,\pi_{vptS})$ represents a vector of model probabilities that voter $v$ scores performer $p$ exactly $s \in \mathcal{S}$ points on occasion $t$. 

As mentioned earlier, in addition to the main outcome, we observe some covariates defined at different levels. Formally, we define:
\begin{itemize}
\item The year in which the contest is held as $x_{1t}$. To simplify the interpretation we actually include in the model the derived variable representing the difference between the year under consideration and the first year in the series, $x^*_{1t}=x_{1t}-1998$. Including this covariate in the model is helpful in accounting for external factors, specific to the particular contest, that may have affected the observed scores;
\item The language in which a song is sung as $x_{2pt}$. This can take on the values 1 = \textit{English}, 2 = \textit{own}, 3 = \textit{mixed} (\textit{i.e.} a combination of two or more languages);
\item The type of performance as $x_{3pt}$. This can take on the values 1 = \textit{Group}, 2 = \textit{Female solo artist}, or 3 = \textit{Male solo artist}.
\end{itemize}
Since $x_{2pt}$ and $x_{3pt}$ are categorical variables, we define suitable dummies $x^{(c)}_{lpt}$ for $l=2,3$ and $c=1,\ldots C_l$, taking value 1 if $x_{lpt}=c$ and 0 otherwise. Thus, $C_2=3$ and $C_3=3$.

Following standard notation in ordinal regression \citep{McCullagh:1980,Congdon:2007,Jackman:2009}, we model the cumulative probabilities $\eta_{vpts} := \Pr(y_{vpt} \leq s)$ as
\begin{eqnarray}\label{Ordinal}
\mbox{logit}(\eta_{vpts}) = \lambda_s - \mu_{vpt},
\end{eqnarray}
with the obvious implication that $\pi_{vpt1} = \eta_{vpt1}$; $\pi_{vpts} = \eta_{vps}-\eta_{vpt(s-1)}$, for $s=2,\ldots,S-1$; and $\pi_{vptS} = 1-\eta_{vptS}$. Here, $\boldsymbol\lambda=(\lambda_1,\ldots,\lambda_S)$ is a set of random cutoff points for the latent continuous outcome associated with the observed categorical variable. In order to respect the ordering constraint implicit in the ordinal structure of the data, we model
\begin{eqnarray*}
\lambda_1 & \sim & \mbox{Normal}(0,\sigma^2_\lambda)\mathbb{I}(-\infty,\lambda_2), \\
\lambda_2 & \sim & \mbox{Normal}(0,\sigma^2_\lambda)\mathbb{I}(\lambda_1,\lambda_3), \\
& \ldots & \\
\lambda_{S-1} & \sim & \mbox{Normal}(0,\sigma^2_\lambda)\mathbb{I}(\lambda_{S-2},\lambda_S), \\
\lambda_S & \sim & \mbox{Normal}(0,\sigma^2_\lambda)\mathbb{I}(\lambda_{S-1},\infty).
\end{eqnarray*} 
Assuming a large variance with respect to the scale in which the variables $\lambda_s$ are defined (\textit{e.g.} $\sigma^2_\lambda=10$) effectively ensures that the strengh of the prior is not overwhelming in comparison to the evidence provided by the data. In addition, the linear predictor $\mu_{vpt}$ is defined as a function of the relevant covariates
\begin{eqnarray}
\mu_{vpt} = \beta_1 x^*_{1t} + \sum_{c=2}^{C_2} \beta_{2c} x^{(c)}_{2pt} +\sum_{c=2}^{C_3} \beta_{3c} x^{(c)}_{3t} + \alpha_{vp}.  \label{linpred}
\end{eqnarray}
The vector of unstructured (fixed) coefficients is defined as $\boldsymbol\beta=(\beta_1,\boldsymbol\beta_2,\boldsymbol\beta_3)$, with $\boldsymbol\beta_2=(\beta_{22},\beta_{23})$ and $\boldsymbol\beta_3=(\beta_{32},\beta_{33})$. The elements in $\boldsymbol\beta$ measure the impact of the covariates on the probability that, on occasion $t$, performer $p$ receives a vote in $\mathcal{S}$ from voter $v$. We consider as reference categories the values \textit{English} for $x^{(c)}_{2pt}$, and \textit{Group} for $x^{(c)}_{3pt}$. As is clear from (\ref{Ordinal}), the model is set up under a proportional odds assumption, \textit{i.e.} that the effect of the predictors is constant across the ordered categories. The negative sign in (\ref{Ordinal}) helps with the interpretation of the $\boldsymbol\beta$ coefficients: larger coefficients are associated with higher probability of a higher score.  

We specify independent and minimally informative Normal priors for the unstructured coefficients
\[ \boldsymbol\beta \sim \mbox{Normal}(\mathbf{m},\mathbf{Q})\]
where $\mathbf{m}$ is a vector of zeros of length $B=\left(1+\sum_{l=2}^3C_l\right)=5$ (\textit{i.e.} the length of the vector $\boldsymbol\beta$) and $\mathbf{Q}=q^2 \mathbf{I}_B$ is a $(B\times B)$ diagonal covariance matrix with $q=10^4$.

\subsection{Modelling the structured effect $\alpha_{vp}$}
The coefficient $\alpha_{vp}$ is the parameter of main interest in our analysis and it represents a structured (random) effect, accounting for clustering at the voter-performer level, which is implied by the fact that we observe repeated instances of the voting pattern from country $v$ towards country $p$, over the years. 

We use a formulation
\[ \alpha_{vp} \sim \mbox{Normal}(\theta_{vp},\sigma^2_\alpha), \]
where the mean is specified as
\begin{eqnarray}
\theta_{vp} = \gamma + \delta_{R_{v}p} + \psi w_{vp} + \phi z_{vp}\mathbb{I}(z_{vp}). \label{theta}
\end{eqnarray}
Here, the coefficient $\gamma$ represents the overall intercept; the covariate $w_{vp}$ takes value 1 if countries $v$ and $p$ share a geographic border and 0 otherwise; and the covariate $z_{vp}$ represents an estimate of the migration intensity from country $v$ to country $p$. Thus, $\psi$ is the ``geographic'' effect and $\phi$ is the ``migration'' effect. Notice that, by design, if there is no recorded migration from $v$ to $p$ we automatically set this effect to 0. 

In addition, we assume that voters implicitly cluster in $K$ ``regions''; this accounts for similarities in voters' propensity towards the performers, over and above the geographic and migratory aspects defined above. For example, because of ``cultutral'' proximity, countries in the Former Soviet bloc may have the same attitude towards one of the performers $p$, regardless of whether they are close geographically or the amount of migration from $p$. For each voter we define a latent categorical variable $R_v$ which can take values $1,2,\ldots,K$ (for a fixed upper bound $K$), \textit{i.e.} $R_v \sim \mbox{Categorical}(\boldsymbol\zeta)$, where $\boldsymbol\zeta=(\zeta_1,\ldots,\zeta_K)$ is the vector of probabilities that each voter belongs in each of the clusters. We use a minimally informative Dirichlet prior on $\boldsymbol\zeta$. Consequently, the coefficients $\delta_{kp}$ (for $k=1,\ldots,K$) represent a set of structured common residual for each combination of macro-area and performer, which we use to describe the ``cultural'' effect.

We model the parameters in the linear predictor for $\theta_{vp}$ using the following specification: $\gamma$, $\psi$ and $\phi$ are given independent minimally informative Normal distributions (centred on 0 and with large variance), while $\delta_{kp}$ are given an exchangeable structure
\begin{eqnarray*}
\delta_{kp} \sim \mbox{Normal}(0,\sigma^2_\delta).
\end{eqnarray*}
The two structured variances are given independent minimally informative prior on the log standard deviation scale
\begin{eqnarray*}
\log(\sigma_\alpha), \log(\sigma_\delta) \stackrel{iid}{\sim} \mbox{Uniform}(-3,3).
\end{eqnarray*}
Since the priors for both $\sigma_\alpha$ and $\sigma_\delta$ are defined on the log scale, a range of $(-3,3)$ is in fact reasonably large and thus these distributions do not imply strict prior constraints on the range of the variability. Sensitivity analyses upon varying the scale of the Uniform distributions have confirmed that the results are generally insensitive to this aspect of the modelling.

The coefficients $\alpha_{vp}$ have an interesting interpretation: consider two voters $v_1$ and $v_2$ and one performer $p$; for each fixed score $s$, $\alpha_{v_1p}$ and $\alpha_{v_2p}$ determine the difference in the estimated probability that either voter would score the performer at most $s$ points, $\eta_{vpts}$, all other covariates being equal (notice that, in our model, none of them depend on the voter anyway). In fact, it easily follows from (\ref{Ordinal}) and (\ref{linpred}) that, if $\alpha_{v_1p}>\alpha_{v_2p}$, for any possible score $s$ the chance that $v_1$ scores $p$ more than $s$ points is greater than the chance that $v_2$ will.

In this sense, we can use the coefficients $\alpha_{vp}$ to quantify the presence of ``favoritism'' or ``discrimination'' between specific countries. Estimated values of $\alpha_{vp}$ substantially below 0 indicate that voter $v$ tends to systematically underscore performer $p$, while values substantially above 0 suggest a systematic pattern in which $v$ scores $p$ higher votes than other voters. Of course, we cannot grant a causal interpretation to this analysis: the acts of favouritism or discrimination imply some deliberate intervention, which we are not able to capture from our data. Nevertheless, we can interpret the estimated values for $\alpha_{vp}$ as at least \textit{indicative} of the underlying voting patterns.

\subsection{Estimation procedure}
The posterior distributions for the parameters of interest are obtained through a MCMC simulation, implemented in \texttt{WinBUGS} \citep{WinBUGS:1996,BUGS:2012}, which we have integrated within \texttt{R} using the library \texttt{R2WinBUGS} \citep{R2WB} $-$ the code to run the model is available on request. Since the model is relatively computationally intensive, we used the \texttt{R} library \texttt{snowfall}, which allows multicore computation. The results are based on 2 chains. For each, we considered 11\,000 iterations following 1\,000 burnin; in addition we thinned the chains selecting one iteration every 20. 

Convergence to the relevant posterior distributions has been checked visually through traceplots and density plots, as well as analytically through the Gelman Rubin diagnostic \citep{Gelman:1992} and the analysis of autocorrelation and the effective sample size.

\section{Results}\label{Results}
We tested three different versions of our model, upon varying the number of possible ``regions'' in which the voters can cluster. We tried values of $K=3,4,5$ and compared the resulting models using the Deviance Information Criterion (DIC, Spiegelhalter et al., 2002). The preferred model is the one with 4 regions (DIC = 36\,832, while the models with 3 and 5 components have DIC = 36\,868 and DIC = 36\,844, respectively). \nocite{Spiegelhalter:2002}

\begin{figure}[!h]
\centering
\includegraphics[scale=.52,angle=-90]{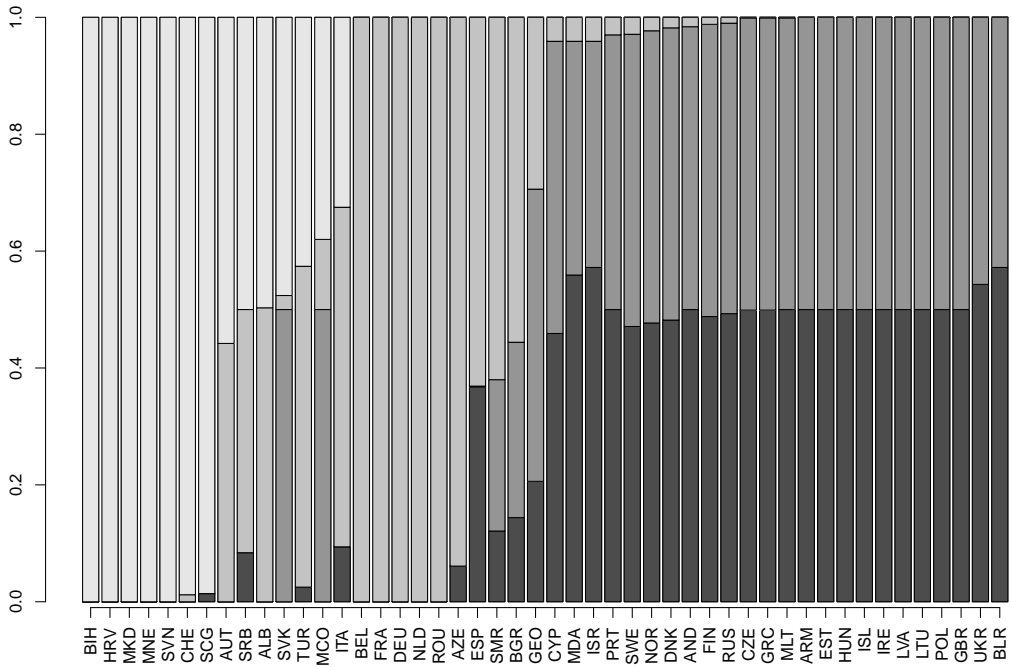}
\caption{Posterior probability that each voter belongs in one of the 4 regions. The lightest shade of grey indicates the cluster (region) labelled as ``1'', while increasingly darker shades of grey indicate regions ``2'', ``3'' and ``4'', respectively. Countries are labelled using their ISO country code, as follows: BIH=Bosnia and Herzegovina; HRV=Croatia; MKD=Macedonia; MNE=Montenegro; SVN=Slovenia; CHE=Switzerland; SCG=Serbia and Montenegro; AUT=Austria; SRB=Serbia; ALB=Albania; SVK=Slovakia; TUR=Turkey; MCO=Monaco; ITA=Italy; BEL=Belgium; FRA=France; DEU=Germany; NLD=Netherlands; ROU=Romania; AZE=Azerbaijan; ESP=Spain; SMR=San Marino; BGR=Bulgaria; GEO=Georgia; CYP=Cyprus; MDA=Moldova; ISR=Israel; PRT=Portugal; SWE=Sweden; NOR=Norway; DNK=Denmark; AND=Andorra; FIN=Finland; RUS=Russia; CZE=Czech Republic; GRC=Greece; MLT=Malta; ARM=Armenia; EST=Estonia; HUN=Hungary; ISL=Iceland; IRE=Ireland; LVA=Latvia; LTU=Lithuania; POL=Poland; GBR=United Kingdom; UKR=Ukraine; BLR=Belarus }\label{Barplot_region}
\end{figure}

Figure \ref{Barplot_region} shows the posterior probability that each voter belongs in one of the 4 clusters. We have labelled them as ``1'', ``2'', ``3'' and ``4'' (they appear in Figure \ref{Barplot_region} in increasing shades of grey, \textit{i.e.} region ``1'' is the lightest and region ``4'' the darkest).

Countries in the former Yugoslavia (notice that because of political changes occurred during the period considered, Serbia and Montenegro are present both as a single country and separately) are clearly clustered in region ``1'', where also Switzerland and Austria tend to feature. This can be explained by their close geographical proximity with the Balkans as well as possible migrations after the 1990's war. 
Region ``2'' is mainly composed by voters in central and southern Europe, but curiously countries such as Bulgaria and Romania tend to cluster in this group, too. This is possibly due to illegal migrations, especially from Romania towards countries such as Spain or Italy. In addition, countries such as Turkey and Albania show a large propensity of clustering in this region. Regions ``3'' and ``4'' show a lower degree of separation and tend to include countries in the Former Soviet bloc (mainly in region ``4'') and countries in northern Europe (specifically Scandinavian countries as well as the UK and Ireland). This result is overall in line with the findings of \citet{Yair:1995}.

Table \ref{FixedEffects} shows the posterior mean and 95\% credibility interval for the unstructured effects from the regression model. We re-iterate that these are not the main interest of the analysis and are included in the model primarily to adjust for potential unbalance in the background characteristics of every voting occasion. Nevertheless, it is possible to see that the analysis of $\boldsymbol\beta_{2}$ suggests that performers singing in their own language are generally scored lower than those singing in English. Also from the results for the coefficients in $\boldsymbol\beta_{3}$ it appears that female solo artists tend to get higher scores than group performances. Both the unstructured geographic effect and migration effect seem to be positively associated with higher scores. Performing countries tend to be scored highly by their neighbours and by countries where their population tend to migrate. 

\begin{table}
\centering
\begin{tabular}{lccc}
\hline
Coefficient (variable) & Mean &    \multicolumn{2}{c}{95\% Credible interval} \\
\hline
$\beta_{1}$  \hspace{5pt}(Year)&     --0.034 &     --0.044 &     --0.023 \\
$\beta_{22}$ \hspace{2pt}(Mixed language)$^a$ &  \textcolor{white}{--}0.062 &     --0.066 &       \textcolor{white}{--}0.194 \\
$\beta_{32}$ \hspace{2pt}(Own language)$^a$ &     --0.131 &     --0.255 &     --0.010 \\
$\beta_{32}$ \hspace{2pt}(Solo female artist)$^b$ &       \textcolor{white}{--}0.232 &       \textcolor{white}{--}0.131 &       \textcolor{white}{--}0.328 \\
$\beta_{33}$ \hspace{2pt}(Solo male artist)$^b$ &     --0.067 &     --0.170 &       \textcolor{white}{--}0.034 \\
$\psi$ \hspace{8pt}(Geographic effect) & \textcolor{white}{--}1.210 & \textcolor{white}{--}0.996 & \textcolor{white}{-}1.430 \\
$\phi$ \hspace{9pt}(Migration effect) & \textcolor{white}{--}0.101 & \textcolor{white}{--}0.076 & \textcolor{white}{--}0.126 \\
\hline
\multicolumn{4}{l}{$^a$ Reference: English}\\
\multicolumn{4}{l}{$^b$ Reference: Group artist}\\
\hline
\end{tabular}
\caption{\label{FixedEffects} Summary of the posterior distributions for the ustructured effects of the regression model}
\end{table}

More interestingly, for each pair $(v,p)$, we can analyse the structured effects $\alpha_{vp}$, describing the systematic components in the voting patterns. In order to make the results comparable on the same scale, we standardised them, \textit{i.e.}\ we centred them around the observed grand mean and divided by the observed overall standard deviation, \textit{e.g.}
\[ \alpha^*_{vp} = \frac{\alpha_{vp} - \bar\alpha}{s_\alpha}, \]
with
\[\bar\alpha = \sum_{v=1}^V \sum_{p=1}^P \frac{\alpha_{vp}}{H} \qquad \mbox{ and } \qquad s_\alpha = \sum_{v=1}^V \sum_{p=1}^P \frac{(\alpha_{vp}-\bar\alpha)^2}{H-1}. \]

Standardisation of the coefficients makes it easier to select some arbitrary thresholds above or below which the effect can be considered to be ``substantial'', therefore indicating the presence of bias. Since, as confirmed by the analysis of the posterior distributions (not shown), the $\alpha_{vp}$ are reasonably normally distributed, we consider a threshold of $\pm 1.96$. Thus, values of $\alpha^*_{vp} > 1.96$ suggest positive bias (``favouritism'') from $v$ to $p$, while values of $\alpha^*_{vp} < -1.96$ are indicative of negative bias (``discrimination'') from $v$ against $p$. 

\begin{figure}[!h]
\centering
\includegraphics[scale=.60]{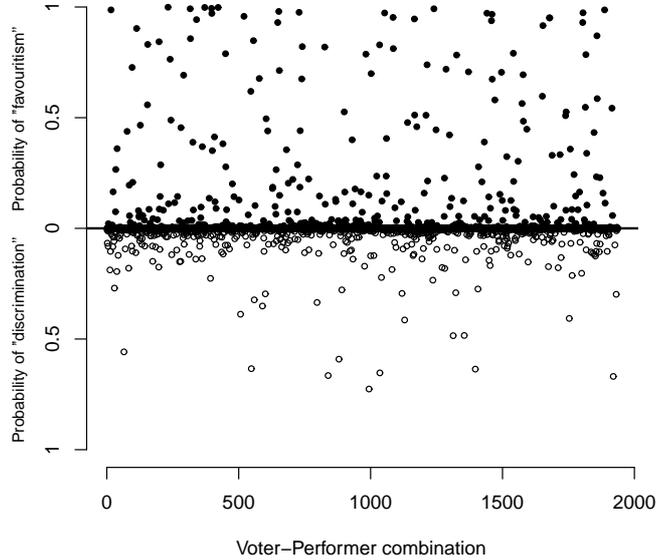}
\caption{Posterior probability that the standardised structured effects exceed the thresholds $\pm 1.96$: The open circles $\circ$ in the bottom part of the graph indicate $\Pr(\alpha^*_{vp} < -1.96 \mid \mathbf{y})$, which represents discrimination from $v$ to $p$, while the solid circles $\bullet$ in the top part of the graph indicate $\Pr(\alpha^*_{vp} > 1.96 \mid \mathbf{y})$, which describes positive bias from $v$ to $p$}\label{Thresholds}
\end{figure}

The analysis of the entire distributions for the $\alpha^*_{vp}$ confirms the absence of clear negative bias throughout the set of voters and performers. In other words, no evidence is found to support the hypothesis that one of the voters \textit{systematically} ``discriminates'' against one of the performers. On the other hand, some patterns of positive bias do emerge from the analysis. This is evident from Figure \ref{Thresholds}: for each of the $H$ voter-performer combinations, the solid circles represent the posterior probability of a positive bias: $\Pr(\alpha^*_{vp} > 1.96 \mid \mathbf{y})$, while the open circles are the posterior probability of a negative bias: $\Pr(\alpha^*_{vp} < -1.96 \mid \mathbf{y})$. As is possible to see, the latter never exceeds 0.75.

\begin{figure}[!h]
\centering
\subfigure[Sweden]{\includegraphics[scale=.32]{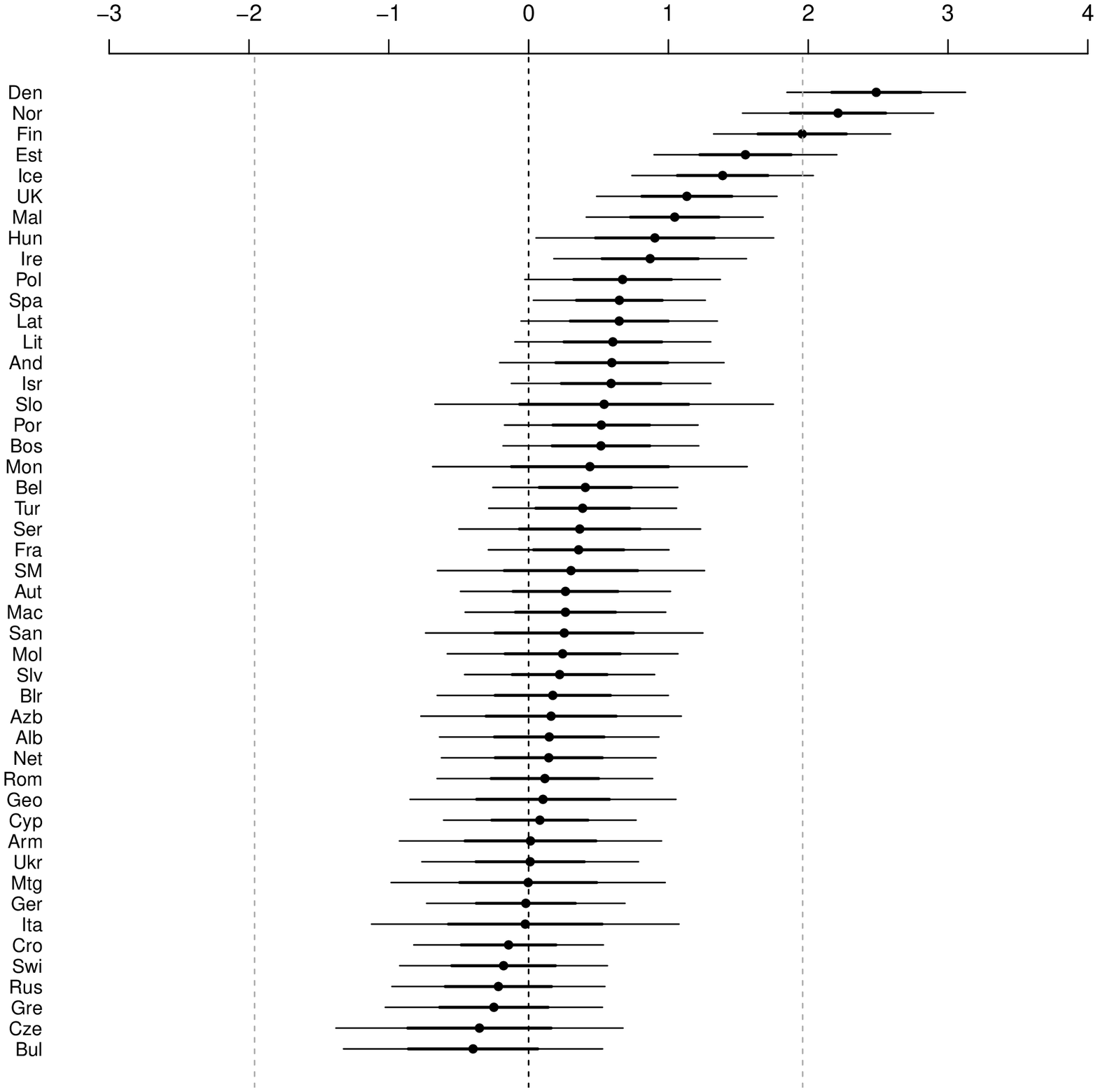}}
\subfigure[Greece]{\includegraphics[scale=.32]{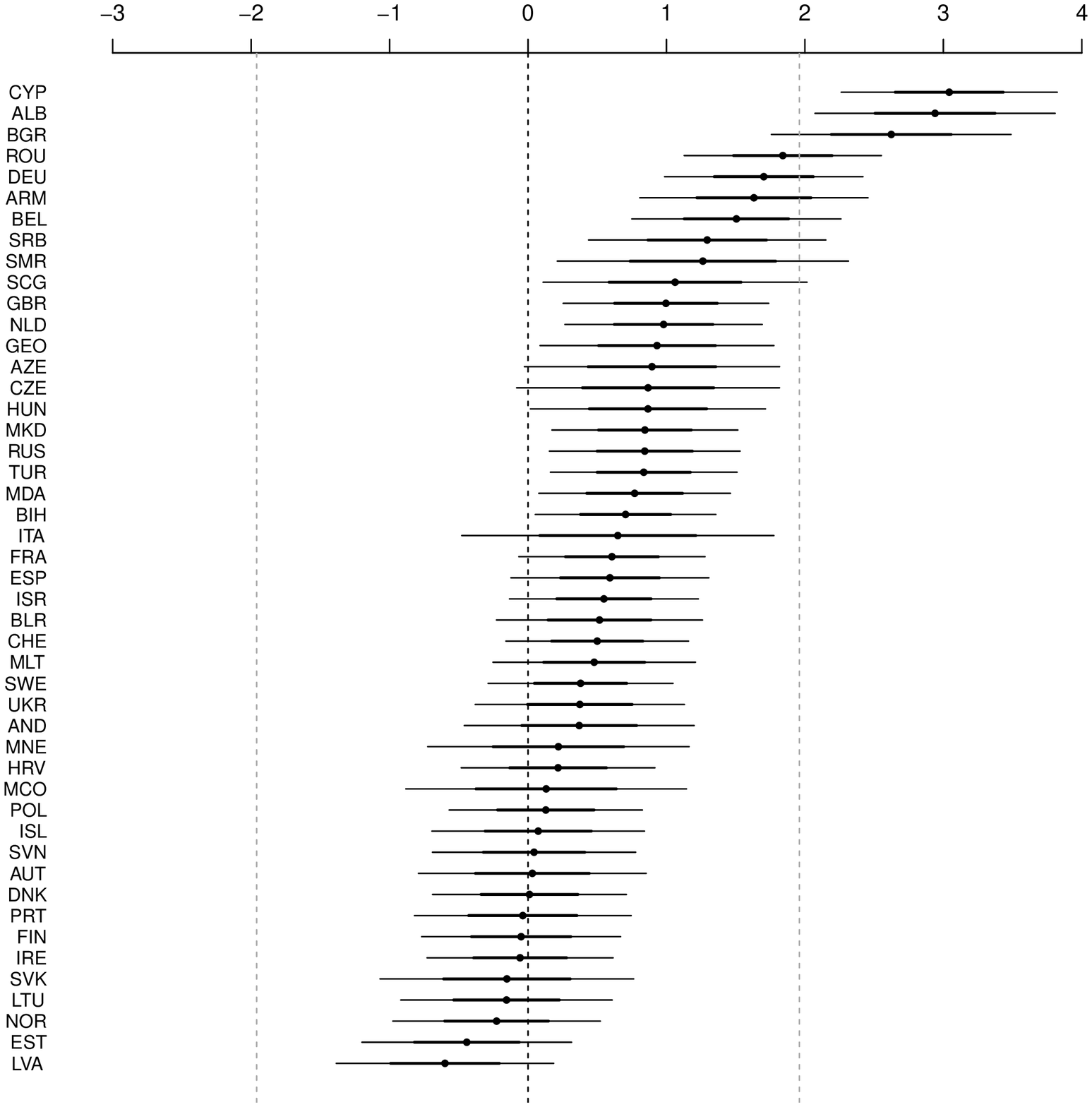}}
\subfigure[Albania]{\includegraphics[scale=.32]{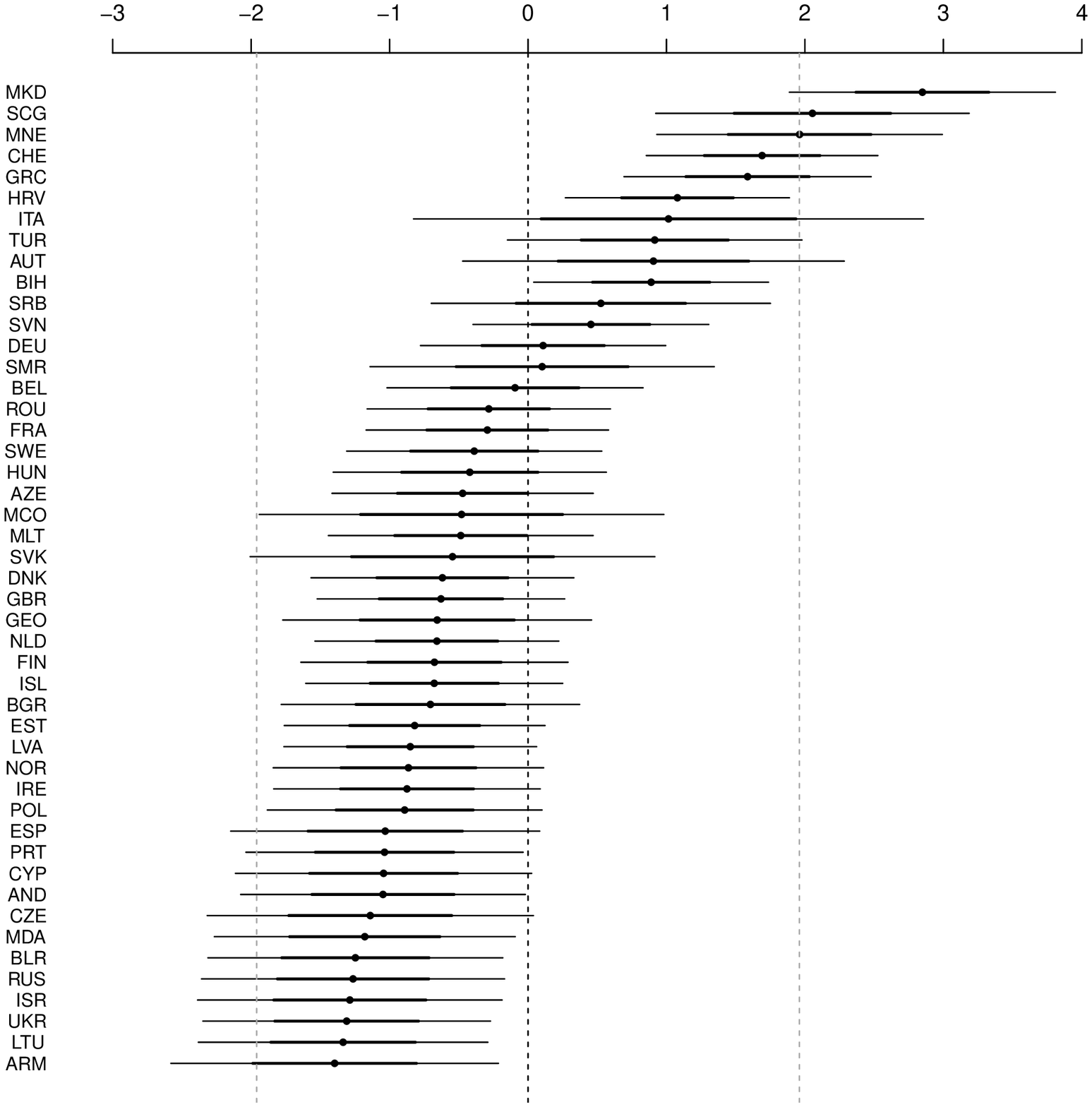}}
\subfigure[Turkey]{\includegraphics[scale=.32]{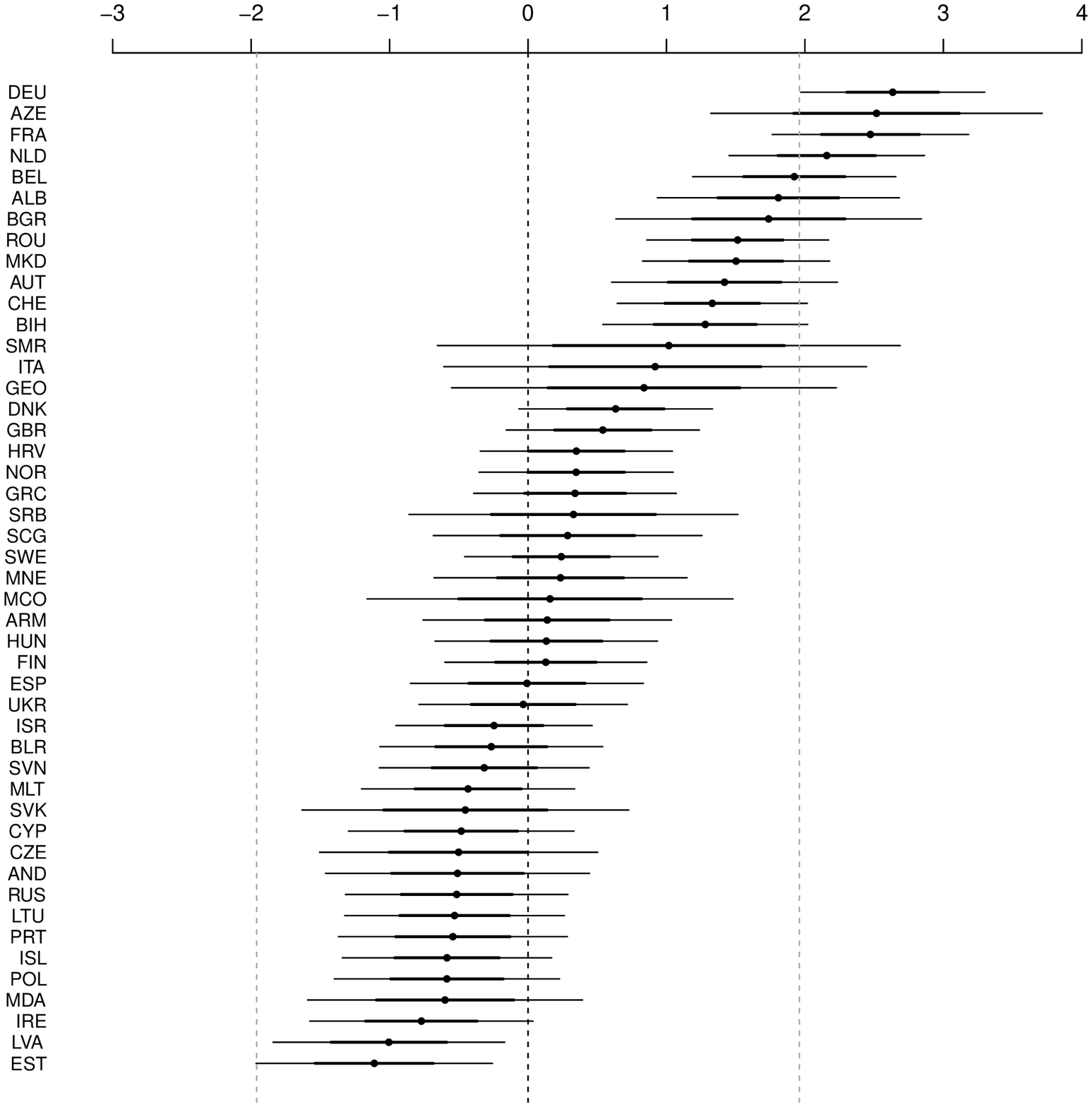}}
\caption{The dots represent the posterior means of the structured effects $\alpha^*_{vp}$, indicating for each voter $v$ the propensity to vote for performer $p$: United Kingdom (top-left corner), Greece (top-right), Albania (bottom-left) and Turkey (bottom-right). Dark and light lines indicate the 50\% and 95\% credibility intervals, respectively}\label{Coefplots}
\end{figure}

Figure \ref{Coefplots} shows a representation of the posterior distributions of the $\alpha^*_{vp}$ for four selected performers. The wide variability in the range of the distributions is driven by the fact that the data are unbalanced, \textit{i.e.} not all the countries compete in every year under investigation. Therefore, the estimation of the coefficient for some of the combinations of voter-performer may be based on only a few instances, thus inducing wide variability --- \textit{e.g.} Italy in Figure \ref{Coefplots}(c).

The voting patterns towards Sweden show a clear absence of any systematic negative bias, since no distribution is entirely below zero, let alone the threshold of $-1.96$. Many of the countries that are closely related to Sweden either geographically or culturally (most notably, Denmark, Norway and Finland) are associated with higher propensity to score the Swedish act higher points. The distribution for Denmark is nearly all above the threshold of $1.96$, indicating a potential positive bias. 

The analysis for other performers show also interesting behaviours: for example, Greece seems to be substantially favoured by its close neighbours Cyprus (for which similarity is geographic as well as cultural) and Albania. Moreover, there is a very large set of voters for which the entire distribution of $\alpha^*_{vp}$ is completely above 0, while no distribution is completely below 0. This indicates a general positive attitude towards Greece, which may be fostered by widespread migrations across Europe.

At the other end of the spectrum, the voting patterns towards Albania are characterised by a large number of voters showing a distribution entirely below 0. While none exceeds the ``discrimination threshold'' of $-1.96$, this seems to suggest very low popularity among the voters. Neighbouring countries such as Macedonia and Montenegro have higher propensities to vote for Albania, but these are not substantial (\textit{i.e.}\ they are never greater than the ``favouritism threshold''	 of 1.96).

Finally, Turkey seems to be substantially favoured in Germany --- possibly due to the large number of Turkish migrants living in (and potentially tele-voting from) Germany. A few other distributions are entirely above 0; for many of those the same migration arguments can be brought forward, while for Azerbaijan there probably are cultural similarities that increase the propensity to vote for Turkey.

\section{Discussion}\label{Discussion}
In this paper we have tried to seek empirical evidence of systematic bias in the Eurovision contest voting. In particular, we have tried to disentangle the two possible extreme behaviours of negative (which may be indicative of ``discrimination'') and positive bias (which possibly suggests ``favouritism''), defined in terms of tail probabilities.

We have used a hierarchical structure to account for correlation induced in repeated instances of the same voter-performer pattern, which occur over time. This by necessity causes shrinkage in the estimations; on the one hand, this potentially limits the ability to identify extreme behaviours. However, on the other hand, because in some cases the sample size observed for a given combination of voter-performer is very small, the hierarchical structure is necessary to avoid unstable estimates for the propensity to vote. In addition, shrinkage is likely to occur on both ends of the distributions; in our results, we are able to identify some examples of ``favouritism'', but no real ``discrimination'' occurs (according to our criteria). Thus, it is reasonable to assume that shrinkage does not impact dramatically on our ability to detect bias.

After having considered some potentially unbalancing factors, we have structured the propensity to vote for a given performer as a function of several components, designed to capture geographic, population movements and cultural effects. The latter has been obtained through a clustering model of the voters embedded in the Bayesian formulation and based on the assumption that voters in the same cultural cluster tend to share similar attitudes towards a given performer. The resulting allocation of voters to the clusters is often consistent with prior expectation about geographical and historical circumstances (\textit{e.g.} the countries in the Former Yugoslavia tend to clearly cluster together). However, because the procedure is mainly data-driven, we gain in flexibility, for example with respect to conditionally autoregressive structures, in modelling spatial correlation.

A related point consists the number of ``regions'' that we have used in the clustering procedure. For simplicity, we have assumed that this was fixed, although we have tested several alternatives to capture the heterogeneity within European nations (for example, to acknowledge the presence of at least four distinct macro-areas: the Former Soviet bloc, Former Yugoslavia, Scandinavia and the rest of Europe).

In conclusion, the findings from our model seem to suggest that no real negative bias emerges in the tele-voting --- in fact, no substantial negative bias occurs across all the combinations of voters-performers. In some cases (and in accordance with previous findings in the literature), we found moderate to substantial positive bias, which could be explained by strong ``cultural'' similarites, \textit{e.g.} due to commonalities in language and history, and to a lesser extent to geographical proximity and migrations. Our formulation highlights the power of Bayesian hierarchical models in dealing with complex data, allowing to properly account for the underlying correlation among the observed data and, possibly, at the higher levels of the assumed structure. 

\bibliographystyle{chicago}
\bibliography{eurofestival}

\end{document}